\documentclass[referee]{raa}            % referee version: for submission

\usepackage{graphicx,times}             %for PS/EPS graphics inclusion, new
\usepackage{natbib}
\usepackage{amssymb,amsmath}
\bibpunct{(}{)}{;}{a}{}{,}

\usepackage[a4paper=true,dvipdfm=true,pagebackref=true]{hyperref}
\hypersetup{colorlinks = true, linkcolor = green, anchorcolor = red, citecolor = blue, filecolor = red, pagecolor = red, urlcolor = red}

\usepackage{lscape}

\begin{document}

   \title{A study on the dynamic spectral indices for SEP events on 2000 July 14 and 2005 January 20}

 \volnopage{ {\bf 20XX} Vol.\ {\bf X} No. {\bf XX}, 000--000}
   \setcounter{page}{1}

   \author{Ming-Xian Zhao\inst{1}, Gui-Ming Le\inst{1}
   }

   \institute{Key Laboratory of Space Weather,
              National Center for Space Weather,
              China Meteorological Administration,
              Beijing, 100081, China; ({\it Legm@cma.gov.cn})\\
\vs \no
   {\small Received 20** June **; accepted 20** July **}
}

\abstract{We have studied the dynamic proton spectra for the two solar energetic particle (SEP) events on 2000 July 14 (hereafter GLE59) and 2005 January 20 (hereafter GLE69). The source locations of GLE59 and GLE69 are N22W07 and N12W58 respectively. Proton fluxes $>$30 MeV have been used to compute the dynamic spectral indices of the two SEP events. The results show that spectral indices of the two SEP events increased more swiftly at early times, suggesting that the proton fluxes $>$30 MeV might be accelerated particularly by the concurrent flares at early times for the two SEP events. For the GLE69 with source location at N12W58, both flare site and shock nose are well connected with the Earth at the earliest time. However, only the particles accelerated by the shock driven by eastern flank of the CME can propagate along the interplanetary magnetic field line to the Earth after the flare. For the GLE59 with source location at N22W07, only the particles accelerated by the shock driven by western flank of the associated CME can reach the Earth after the flare. Results show that there was slightly more than one hour during which the proton spectra for GLE69 are softer than that for GLE59 after the flares, suggesting that the shock driven by eastern flank of the CME associated with GLE69 is weaker than the shock driven by the western flank of the CME associated with GLE59. The results support that quasi-perpendicular shock has stronger potential in accelerating particles than the quasi-parallel shock. The results also suggest that only a small part of the shock driven by western flank of the CME associated with the GLE59 is quasi-perpendicular.
\keywords{Sun: flares --- Sun: coronal mass ejections (CMEs) --- Sun: particle emission --- Sun: solar-terrestrial relations
}
}

   \authorrunning{M.-X. Zhao \& G.-M. Le}            %author_head in even pages
   \titlerunning{Dynamic protons energy spectra comparison}  % title_head in odd pages
   \maketitle

%________________________________________________ sections below
%
\section{Introduction}           %% first-level sections will be auto-capitalized
\label{sect:intro}

Large gradual solar energetic particle (SEP) events are often accompanied by both flare and fast coronal mass ejection (CME). Both the flare and CME-driven shock may contribute to the productions of SEPs. However, whether a gradual flare can accelerate protons to high energy and even to relativistic energy is still an open question. For example, \cite{Reames+1999} suggested that only CME-driven shock can accelerate protons to high energy in large gradual SEP events. However, some researchers argued that flares may dominate in the acceleration of particles at the early phase of large gradual SEP events (e.g., \citealt{Cane+etal+2003}). When relativistic solar protons (RSPs) reach the atmosphere of the Earth, the interaction between the RSPs and the particles of the Earth's atmosphere causes the atmospheric cascade.

At times only a small fraction of the RSPs accelerated to the energy of $\ge$1 GeV generates cascades in the atmosphere sufficiently (see \citealt{Mewaldt+etal+2012, Wu+Qin+2018, Firoz+etal+2019b}). Such RSPs are termed as ground level enhancements (GLEs) registered by neutron monitors on the Earth. \cite{Firoz+etal+2010} proposed that a conjunction between CME-driven interplanetary shock and flare may produce GLE, suggesting that the CME alone presumably does not cause GLE. Many case studies show that RSPs including the two large SEP events associated with GLE59 and GLE69 may be accelerated by the concurrent flares (e.g., \citealt{Firoz+etal+2011, Firoz+etal+2012, Grechnev+etal+2008, Klein+etal+2001, Klein+etal+2014, Le+etal+2006, Le+Zhang+2017, Li+etal+2007, McCracken+etal+2008, Masson+Klein+2009, Simnett+2006, Simnett+2007}).

\begin{figure}
   \centering
  \includegraphics[width=7cm, angle=0]{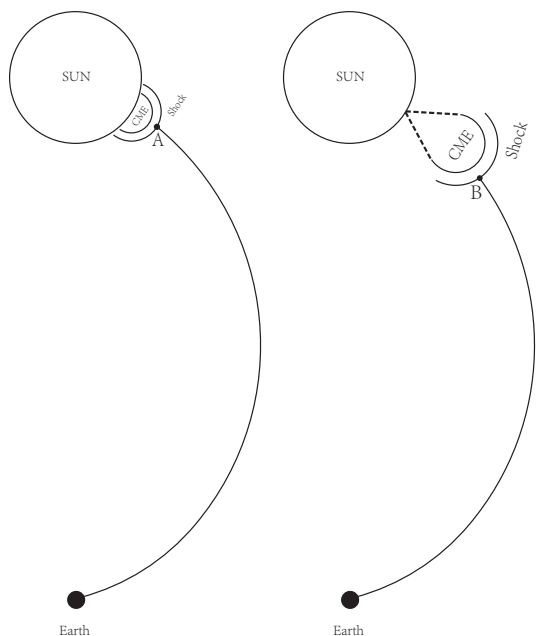}
   \caption{The two positions on the shock surface connect with the Earth at two different moments}
   \label{fig:cobpoint}
\end{figure}

CME-driven shock is naturally a large scale structure, so the particles accelerated by CME-driven shock can be observed in much wider helio-longitudinal area. However, shock strength varies along the shock surface. The shock appears stronger usually at the nose and declines on the flanks of the CMEs. In general, the shock on the eastern flank is quasi-parallel while the shock in the western flank is quasi-perpendicular (\citealt{Reames+1999, Kallenrode+2001, Kahler+2016}). The position on the shock surface connected with the Earth depends on the longitude of the location of CME-driven shock relative to the Earth, and the position changes as the CME moves away from the Sun and propagates in interplanetary space. Here, we give a diagram to illustrate the changes of the positions on the shock surface that connect with the Earth as the CME moves from the Sun, which is shown in Figure \ref{fig:cobpoint}. The Figure \ref{fig:cobpoint} shows that the points A and B, indicating the shock noses, connect the Earth differently. The shock nose marked by the A connects with the Earth at the earliest time, whereas, sometimes later, the shock nose marked by the B located on the eastern flank of the shock surface connects with Earth.

It is yet to be understood about which shock, quasi-perpendicular or quasi-parallel, has stronger potential in accelerating particles. Based on the simulation processes, several researchers opined that the quasi-parallel shock driven by the eastern flank of a CME has stronger potential in accelerating particles than quasi-perpendicular shock driven by western flank of a CME (e.g. \citealt{Li+etal+2003, Li+etal+2005, Zank+etal+2006}). Some other researchers argued that quasi-perpendicular shock driven by western flank of a CME has stronger potential in accelerating particles than quasi-parallel shock driven by eastern flank of a CME (e.g. \citealt{Jokipii+1987, Qin+etal+2018}).

The intensity-time profile of a SEP event depends on the longitude of the source location of SEP event relative to the observer and the interplanetary shock driven by associated CME (\citealt{Cane+etal+1988}). For Geostationary Operational Environmental Satellites (GOES; e.g., \citealt{Aschwanden+and+Freeland+2012}) that observed the proton fluxes used in this study, the intensity-time profile of an SEP event not only depends on the concurrent solar acceleration processes (flare; CME) and the longitude of the source location of SEP event relative to GOES, but also depends on location of interplanetary shock relative to GOES and the intensity of interplanetary shock. Figure \ref{fig:cobpoint} shows the variation of the points on the shock surface connected with the Earth. Because shock intensity at different points on shock surface is different, hence the energy spectral index of particles observed by GOES should change continuously as CME moves away from the Sun and propagates in interplanetary space.

To investigate possible source for the earliest particles accelerated by associated solar flares and check whether perpendicular shock is more effective in accelerating protons than parallel shock, dynamic energy spectral indices of protons for GLE59 and GLE69 are to be computed and compared in this study. The energy spectral indices of double power law for SEP events associated with GLE59 and GLE69 have been determined by a few researchers (\citealt{Mewaldt+etal+2012, Wu+Qin+2018}). However, the double power law is the event integrated differential spectra, which cannot reflect the variation of the energy spectral index with time. This article is arranged as follows. Data analysis is presented in section \ref{sect:data}. Discussion is given in section \ref{sect:discussion}. Summary and conclusion are noted in the final section.

\section{Data analysis}
\label{sect:data}
\subsection{Observations}
\label{sect:obs}
\subsubsection{GLE59 on 2000 July 14}
\label{sect:obs0714}

\begin{landscape}
\begin{figure}
   \centering
   \includegraphics[width=12cm,angle=90]{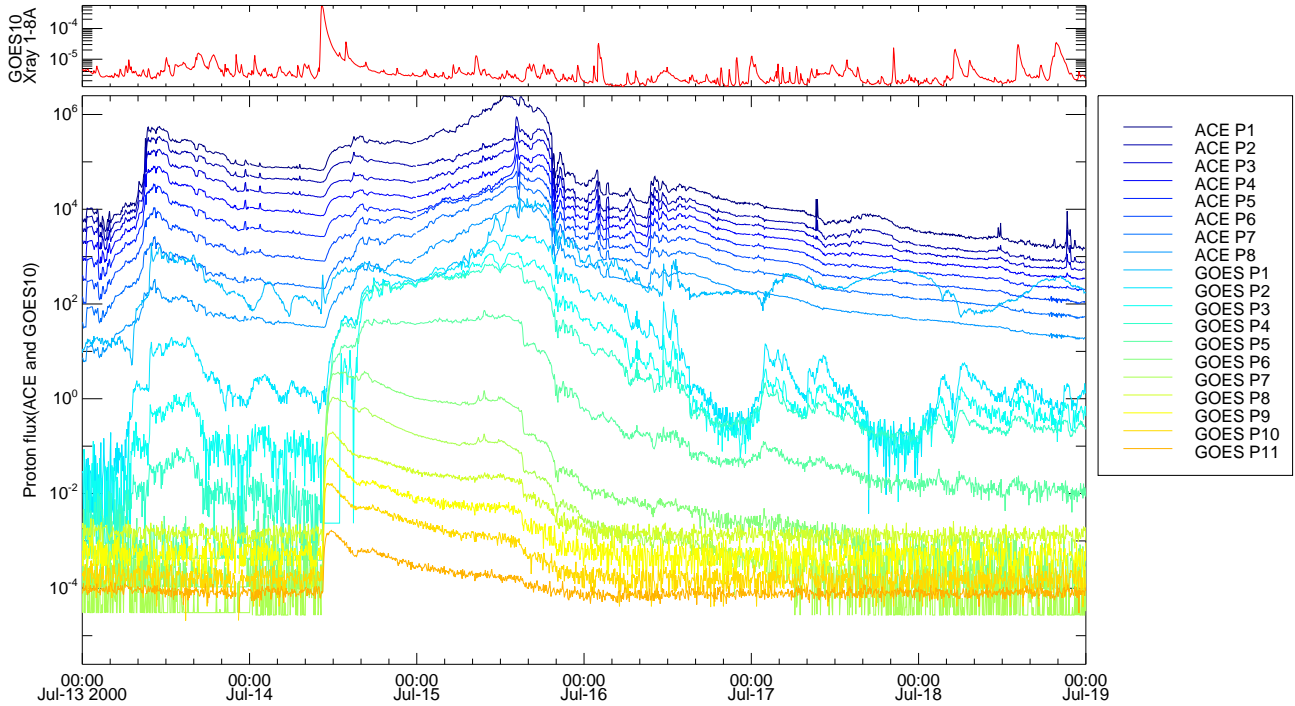}
   \caption{Fluxes of solar soft X-rays and proton particles with different energies during 13-18, July 2000. The upper panel shows the flux of soft X-ray in 1-8~\AA~ observed by GOES. The lower panel shows ACE EPAM/LEMS120 (\citealt{Gold+etal+1998}) ion fluxes with energy (P1)0.047-0.068 MeV, (P2)0.068-0.115 MeV, (P3)0.115-0.195 MeV, (P4)0.195-0.321 MeV, (P5)0.310-0.580 MeV, (P6)0.587-1.060 MeV, (P7)1.060-1.900 MeV and (P8)1.900-4.800 MeV. GOES EPS corrected proton flux with energy (P1)0.6-4.0 MeV, (P2)4.0-9.0 MeV, (P3)9.0-15.0 MeV, (P4)15.0-44.0 MeV, (P5)40.0-80.0, (P6)80.0-165.0 and (P7)165.0-500.0 MeV. GOES HEPAD proton flux with energy (P8)350.0-420 MeV, (P9)420-510 MeV, (P10)510-700 MeV, and (P11)$>$700 MeV. All data are of 5 min resolution.}
   \label{fig:obs0714}
\end{figure}
\end{landscape}

Solar active region(SAR) 9077, which is located at N22W07, produced a X5.7 flare. The flare started at 10:03 UT and peaked at 10:24 UT on 2000 July 14 and then a CME associated with the flare firstly entered Solar and Heliospheric Observatory (SOHO)/Large Angle and Spectrometric Coronagraph (LASCO)-C2 field of view (2.2-6 $R_s$) at 10:54 UT. The projected speed of the CME was 1674 km/s (\url{https://cdaw.gsfc.nasa.gov/CME_list/}; e.g., \citealt{Yashiro+etal+2004}). A large gradual SEP event accompanied with the flare and CME was observed by ACE and GOES 8, which is shown in Figure \ref{fig:obs0714}.

\begin{landscape}
\begin{figure}
   \centering
  \includegraphics[width=12cm,angle=90]{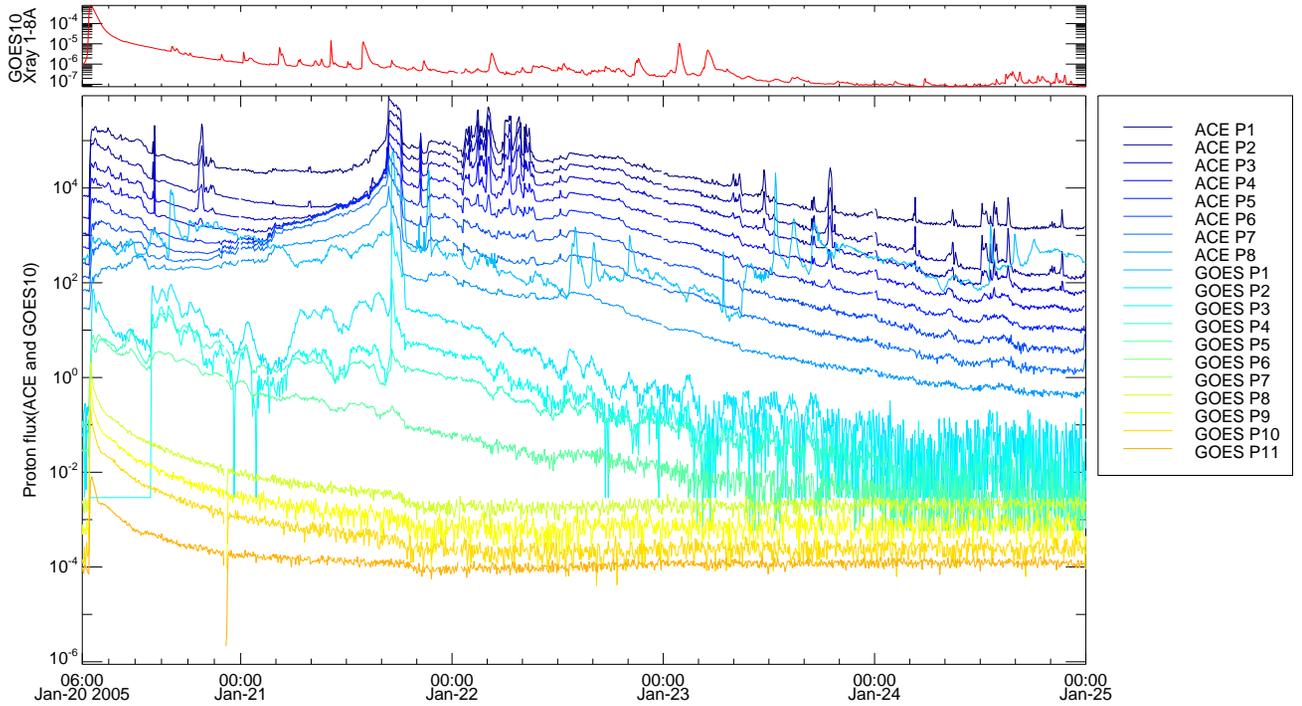}
   \caption{Fluxes of solar soft X-rays and proton particles with different energies during 20-24, January 2005.}
   \label{fig:obs0120}
\end{figure}
\end{landscape}

\subsubsection{GLE69 on 2005 January 20}
\label{sect:obs0120}

SAR 10720 located at N12W58 produced a X7.1 flare. The flare started at 06:36 UT and peaked at 07:01 UT on 2005 January 20. A CME associated with the flare with a projected speed 882 km/s was first observed by SOHO/LASCO C2 at 06:54 UT on 2005 January 20(\url{https://cdaw.gsfc.nasa.gov/CME_list/}; e.g., \citealt{Yashiro+etal+2004}). A large gradual SEP event accompanied with the flare and CME was observed by ACE and GOES 11, which is shown in Figure \ref{fig:obs0120}.

\subsection{Comparison between proton fluxes$>$100 MeV for two SEP events}
\label{sect:compareflux}

\begin{figure}
   \centering
  \includegraphics[width=10cm, angle=0]{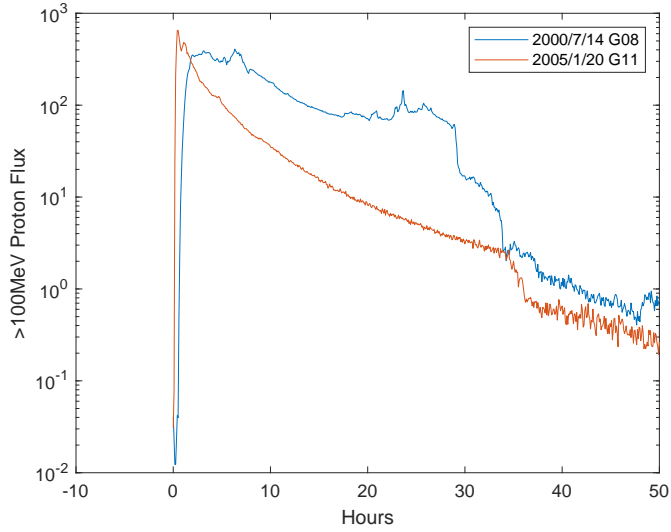}
   \caption{Comparison between fluxes of E$>$100 MeV protons of two major SEP events}
   \label{fig:compareflux}
\end{figure}

The fluxes of protons with different energies usually increase swiftly after the eruptions of the associated flares and CMEs. The flux of energy (E)$>$100 MeV proton usually reached its peak flux no longer after the eruptions of the associated flare and CME, suggesting that the strongest acceleration for E$>$100 MeV proton takes place in the Sun or in the interplanetary space near the Sun. The fluxes of E$>$100 MeV protons for GLE59 and GLE69 are shown in Figure \ref{fig:compareflux}. It is seen that the flux of E$>$100 MeV proton of the GLE69 event reached its peak flux faster than that of GLE59. The peak fluxes of E$>$100 MeV protons of the GLE69 and GLE59 events are 698 pfu and 408 pfu, respectively (\citealt{Le+etal+2016, Le+etal+2017}). [1 proton flux unit (pfu)=$cm^{-2}sr^{-1}s^{-1}$].

It is evident that peak flux of E$>$100 MeV proton of the GLE69 is much stronger than that of the GLE59. However, the flux of E$>$100 MeV proton of the GLE69 decayed much faster than that of the GLE59 after their peak fluxes. The source location of the GLE69 is N12W58, which is well connected with the Earth, because the location is far away from the solar center (\citealt{Swalwell+etal+2017}). However, the source location N22W07 of GLE59 is not well connected with the Earth, because the location is close to the solar center. This may be the reason why the flux of E$>$100 MeV of the GLE69 reached its peak flux faster than that of the GLE59.

The shock nose driven by the GLE69-associated CME is well connected with the Earth at the earliest time, and then the eastern flank of the shock is connected with the Earth and the shock intensity declined gradually as the CME moved away from the Sun. On the contrary, the particles accelerated by the western flank of the shock associated with the GLE59 can reach the Earth and the shock intensity also changed continuously as the CME moves away from the Sun. One can understand from Figure \ref{fig:compareflux} that the flux of E$>$100 MeV proton of the GLE59 is stronger than that of the GLE69 no longer after their peak fluxes, suggesting that the intensity of the western flank shock associated with the GLE59 may be stronger than that of eastern flank shock associated with the GLE69.

\subsection{Comparison of the dynamic energy spectral indices of the two SEP events}
\label{sect:comparespectral}

\subsubsection{Method}
Double power laws were used to study the energy spectra of GLEs that occurred during solar cycle 23. The results showed that the breaking energies for GLE59  and GLE69 are 24.2 MeV and 8.18 MeV respectively (\citealt{Mewaldt+etal+2012}). The breaking energies for both GLEs are lower than 30 MeV. Hence, E$>$30 MeV protons observed by GOES are used to calculate the dynamic energy spectral indices for the two SEP event. $f(E) \propto E^{-\gamma_2}$ is used to calculate the dynamic spectral index of the two SEP events. Time resolution of protons observed by GOESs is 5-minutes. The SEP data observed by GOES 8 and GOES 11 are used to calculate the dynamic energy spectral indices for GLE59 and GLE69 respectively. The start times of two flares are all toggled to zero time. 7 differential channels (channels from P5 to P11, energy ranging from 40 to $>$700 MeV), and 4 integral channels ($>$30, $>$50, $>$60, and $>$100 MeV, described in \citealt{Mewaldt+etal+2005}) observed by GOES are used to calculate energy spectral indices.

\subsubsection{Results}
\begin{figure}
   \centering
  \includegraphics[width=10cm, angle=90]{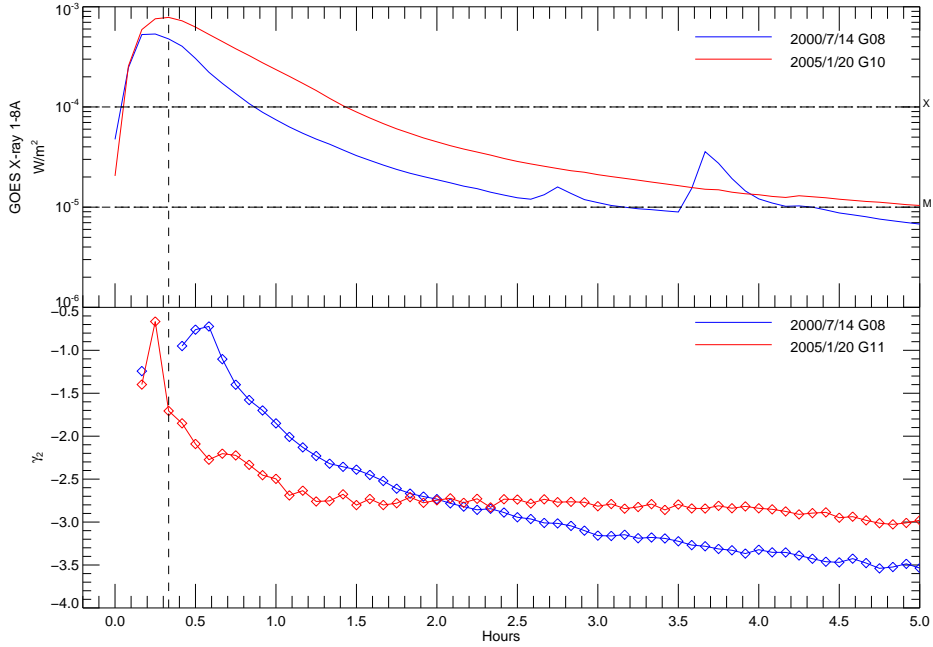}
   \caption{Dynamic energy spectral indices for the two SEP events. The upper and lower panels indicate the flux of solar X-ray in 1-8~\AA~ and dynamic energy spectral indices of the two SEP events respectively. The vertical dashed line indicates the peak time of X7.1 that occurred on 2005 January 20}
   \label{fig:comparespectral}
\end{figure}

The dynamic spectral indices calculated for the GLE59 and GLE69 are shown in Figure \ref{fig:comparespectral}, which exposes that the spectral indices for the two GLEs increased faster and reached peak value promptly. The decay phases of the spectral indices for the two events differ a lot. The decay phase of the spectral index for GLE69 declines much more promptly than that for the GLE59. In fact, the decay phase of the spectral index for GLE59 declines abruptly. In this regard, \cite{Firoz+etal+2019a} observed that the GLE69-associated DH-type II burst ended about 112 min earlier than the flare, implying that the CME shock did not operate over the decay phase of the GLE69 particle event, whereas CME shock operated over the decay phase of the GLE59 particle event.

As mentioned in the earlier (Figure \ref{fig:cobpoint}), the source location for GLE69 is well connected with the Earth that the particles accelerated by the flare and latter reached the CME shock nose can directly propagate to the Earth along the interplanetary magnetic field line at the earliest time. However, only the particles accelerated by eastern flank shock can reached the Earth. For GLE59, only the particles accelerated by western flank shock can reach the Earth. We can also see from Figure \ref{fig:comparespectral} that there was slightly more than 1 hour during which the energy spectral index for GLE59 is higher than that for GLE69, suggesting that western flank shock associated with GLE59 is stronger than eastern flank shock associated with GLE69 during this period. The shock on the eastern flank is quasi-parallel shock, while the shock in the western flank is quasi-perpendicular shock (\citealt{Reames+1999}). In this context, quasi-perpendicular shock is stronger than quasi-parallel shock.

\section{Discussion}
\label{sect:discussion}

The energy spectral indices for GLE69 increased quickly and then reached its peak value at 06:50 UT on 2005 January 20. It is evident that the hardest proton spectrum occurred during the rising phase of the associated flare. Higher energy protons have closer association with the associated flare, while lower energy protons have closer association with associated CME-driven shock (\citealt{Le+Zhang+2017}). In this context, the phenomenon that spectral index increased quickly at early times indicates that E$>$30 MeV protons in the two GLEs should be mainly accelerated by the concurrent flares.

The flux of E$>$100 MeV proton for GLE69 reached its peak flux faster than that for GLE59. The peak flux of E$>$100 MeV proton for GLE69 is much stronger than that for GLE59. The proton spectra for GLE69 is harder than that for GLE59 at early times (Figure \ref{fig:comparespectral}), suggesting that solar eruptions associated with GLE69 have stronger potential to accelerate protons to E$>$100 MeV than that associated with GLE59 at early times. \cite{Gopalswamy+etal+2005} proposed that the speed of the CME associated with GLE69 is 3242 km/s, which is much faster than the CME projected speed 882 km/s. If the speed of the CME associated with GLE69 is really 3242 km/s or even close to this value, the E$>$30 MeV protons may be accelerated by both concurrent flare and CME-driven shock at early times. However, E$>$30 MeV protons may still mainly be accelerated by concurrent flare at early times because proton spectrum became harder at early times.

Figure \ref{fig:comparespectral} shows that there was only slightly more than 1 hour during which the energy spectral index for GLE59 is higher than that for GLE69, suggesting that quasi-perpendicular shock associated with GLE59 is stronger than quasi-parallel shock associated with GLE69, which is consistent with the simulation results obtained by \cite{Qin+etal+2018}. To be noticed, only small part of the western flank shock associated with GLE59 is quasi-perpendicular shock.

The results of the present study support the results obtained by Firoz et al. (2019) that GeV protons are accelerated by concurrent flare. In this context, the result that GeV particles were accelerated by associated flare obtained in the paper of \cite{Zhao+etal+2018} is reasonable. Now the question is for GLE59, how the RSPs accelerated by the associated flare with source location at N22W07 propagated to the Earth? The simulation of the results made by \cite{Zhang+Zhao+2017} showed that if the perpendicular diffusion is about 10\% of what is derived from the random walk of field lines at the rate of supergranular diffusion, particles injected at the compact solar flare site can spread to a wide range of longitude and latitude, very similar to the behavior of particles injected at a large CME shock. The results of present study that E$>$30 MeV proton may be mainly accelerated by concurrent flare associated with GLE59 at early times give an evidence that particles accelerated by associated flare can spread to a wide range of longitude and latitude, very similar to the behavior of particles injected at a large CME shock (\citealt{Zhang+Zhao+2017}).

\section{Summary and Conclusion}
\label{sect:summary}

We have analyzed solar proton fluxes of E$>$30 MeV and studied the spectral indices for the SEP events on 2000 July 14 (GLE59) and 2005 January 20 (GLE69). Important results are summarized as follows.

1. Solar acceleration processes during the GLE69 event has stronger potential to accelerate the protons to GeV energetics than those during the GLE59 event. E$>$30 MeV protons for both the GLE59 and GLE69 seemed to have been accelerated mainly by the flares at early times. Our analysis has been illustrated by the simulation study of \cite{Zhang+Zhao+2017} that the particles injected from the flare site can spread to a wide range of longitude and latitude, which is very similar to the behavior of particles injected at a large CME shock.

The results of our study also support the viewpoints proposed by \cite{Firoz+etal+2019a} that the MeV energetic protons are initiated over the flare initial phases and then accelerated to GeV energetic over the flare prompt phases associated with the coronal shocks manifested in metric-type II burst.

2. The spectral index for GLE59 is higher than that for GLE69 for about 1 hour over the flare decay phases where coronal shocks manifested in DH-type II bursts played much stronger roles for the GLE59 (e.g., see \citealt{Firoz+etal+2019a, Firoz+etal+2019b}). The results suggest that quasi-perpendicular shock associated with GLE59 is stronger than quasi-parallel shock associated with GLE69, and only a small part of the western flank shock associated with GLE59 is quasi-perpendicular shock.

\normalem
\begin{acknowledgements}

We are grateful to SOHO/LASCO,and CDAW, GOES data for making their data available online. This work is funded by the National Natural Science Foundation of China (Grant No. 41674166)

\end{acknowledgements}

\label{lastpage}


\begin{thebibliography}{99}

  \bibitem[Aschwanden \& Freeland(2012)]{Aschwanden+and+Freeland+2012}Aschwanden M. J., \& Freeland S. L., 2012, \apj, 754, 112

  \bibitem[Cane et al.(1988)]{Cane+etal+1988}Cane H. V., Reames D. V., \& Rosenvinge T. T. VONR, 1988, \jgr, 93(A9), 9555

  \bibitem[Cane et al.(2003)]{Cane+etal+2003}Cane H. V., von Rosenvinge T. T., Cohen C. M. S., \& Mewaldt R. A., 2003, \grl, 30(12), 8017

  \bibitem[Firoz et al.(2010)]{Firoz+etal+2010}Firoz K. A., Cho K.-S., Hwang J., et al., 2010, \jgr, 115, A09105

  \bibitem[Firoz et al.(2011)]{Firoz+etal+2011}Firoz K. A., Moon Y.-J., Park S.-H., et al., 2011, \apj, 743, 190

  \bibitem[Firoz et al.(2019a)]{Firoz+etal+2019a}Firoz K. A., Gan W. Q., Li Y. P., et al., 2019a, \apj, 872, 178

  \bibitem[Firoz et al.(2019b)]{Firoz+etal+2019b}Firoz K. A., Gan W. Q., Moon Y. J., et al., 2019b, \apj, 883, 91

  \bibitem[Firoz et al.(2012)]{Firoz+etal+2012}Firoz K. A., Gan W. Q., Moon Y.-J., \& Li, C., 2012, \apj, 758, 119

  \bibitem[Gold et al.(1998)]{Gold+etal+1998}Gold R. E., Krimigis S. M., Hawkins S. E., et al., 1998, \ssr, 86, 541

  \bibitem[Grechnev et al.(2008)]{Grechnev+etal+2008}Grechnev V. V., Kurt V.G., Chertok I. M., et al., 2008, \solphys, 252, 149

  \bibitem[Gopalswamy et al.(2005)]{Gopalswamy+etal+2005}Gopalswamy N., Xie H., Yashiro S., Usoskin I., 2005, in Proceedings of 29th International Cosmic Ray Conference, Vol. 1, ed., B. Sripathi Acharya, S. Gupta, P. Jagadeesan, A. Jain, S. Karthikeyan, S. Morris, S. Tonwar (Tata Institute of Fundamental Research, Mumbai), 169

  \bibitem[Jokipii(1987)]{Jokipii+1987}Jokipii J. R.,1987, \apj, 313, 842

  \bibitem[Kahler(2016)]{Kahler+2016}Kahler S. W., 2016, \apj, 819, 105

  \bibitem[Kallenrode(2001)]{Kallenrode+2001}Kallenrode M.-B., 2001, \jgr, 106, 24989

  \bibitem[Klein et al.(2001)]{Klein+etal+2001}Klein K.-L., Trottet G., Lantos P., \& Delaboudini\'{e}re J.-P., 2001, \aap, 373, 1073

  \bibitem[Klein et al.(2014)]{Klein+etal+2014}Klein K.-L., Masson S., Bouratzis C., et al., 2014, \aap, 572, A4

  \bibitem[Le et al.(2006)]{Le+etal+2006}Le G.-M., Tang Y.-H., \& Han Y.-B., 2006, \chjaa, 6(6), 751

  \bibitem[Le et al.(2016)]{Le+etal+2016}Le G.-M., Li C., Tang Y.-H., et al., 2016, \raa, 16, 14

  \bibitem[Le et al.(2017)]{Le+etal+2017}Le G.-M., Li C., Zhang X.-F., 2017, \raa, 17(7), 73

  \bibitem[Le \& Zhang(2017)]{Le+Zhang+2017}Le G.-M., \& Zhang X.-F., 2017, \raa, 17(12), 123

  \bibitem[Li et al.(2003)]{Li+etal+2003}Li G., Zank G. P., \& Rice W. K. M., 2003, \jgr, 108(A2), 1082

  \bibitem[Li et al.(2005)]{Li+etal+2005}Li G., Zank G. P., \& Rice W. K. M., 2005, \jgr, 110, A06104

  \bibitem[Li et al.(2007)]{Li+etal+2007}Li C., Tang Y. H., Dai Y., Zong W. G., \& Fang C., 2007, \aap, 461, 1115

  \bibitem[Masson \& Klein(2009)]{Masson+Klein+2009}Masson S., Klein K.-L., B\"utikofer R., et al., 2009, \solphys, 257, 305

  \bibitem[McCracken et al.(2008)]{McCracken+etal+2008}McCracken K. G., Moraal H., \& Stoker P. H., 2008, \jgr, 113, A12101

  \bibitem[Mewaldt et al.(2005)]{Mewaldt+etal+2005}Mewaldt R. A., Cohen C. M. S., Labrador A. W., et al., 2005, \jgr, 110, A09S18

  \bibitem[Mewaldt et al.(2012)]{Mewaldt+etal+2012}Mewaldt R. A., Looper M. D., Cohen C. M. S., et al., 2012, \ssr, 171(1-4), 97

  \bibitem[Qin et al.(2018)]{Qin+etal+2018}Qin G., Kong F.-J., \& Zhang L.-H., 2018, \apj, 860:3

  \bibitem[Reames(1999)]{Reames+1999}Reames D. V., 1999, \ssr, 90, 413

  \bibitem[Simnett(2006)]{Simnett+2006}Simnett G. M., 2006, \aap, 445, 715

  \bibitem[Simnett(2007)]{Simnett+2007}Simnett G. M., 2007, \aap, 472, 309

  \bibitem[Swalwell(2017)]{Swalwell+etal+2017}Swalwell B., Dalla S., \& Walsh R.W., 2017, \solphys, 292, 173

  \bibitem[Wu \& Qin(2018)]{Wu+Qin+2018}Wu S.-S., \& Qin G., 2018, \jgr :Space Physics, 123, 76

  \bibitem[Yashiro et al.(2004)]{Yashiro+etal+2004}Yashiro S., Gopalswamy N., Michalek G., et al., 2004, \jgr, 109, A07105

  \bibitem[Zank et al.(2006)]{Zank+etal+2006}Zank G. P., Li G., Florinski V., et al., 2006, \jgr, 111, A06108

  \bibitem[Zhang \& Zhao(2017)]{Zhang+Zhao+2017}Zhang M., \& Zhao L., 2017, \apj, 846, 107

  \bibitem[Zhao et al.(2018)]{Zhao+etal+2018}Zhao M.-X., Le G.-M., \& Chi Y.-T., 2018, \raa, 18(7), 74

\end{thebibliography}
\end{document}